\documentstyle[12pt,epsfig]{article}
\begin{document}
\begin {center}
{\bf {\Large
Resonance dynamics in the coherent $\eta$ meson production in the
$(p,p^\prime)$ reaction on the spin-isospin saturated nucleus} }
\end {center}
\begin {center}
Swapan Das      
\footnote {email: swapand@barc.gov.in} \\
{\it Nuclear Physics Division,
Bhabha Atomic Research Centre  \\
Mumbai-400085, India }
\end {center}

\begin {abstract}
For the forward going proton and $\eta$ meson, the coherent $\eta$ meson
production in the $(p,p^\prime)$ reaction on the spin-isospin saturated
nucleus occurs only due to the $\eta$ meson exchange interaction between
the beam proton and nucleus.
In
this process, the nucleon in the nucleus can be excited to resonances
$N^*$ and the $\eta$ meson in the final state can arise due to
$N^* \to N\eta$.
We
investigate the dynamics of resonances, including nucleon Born terms,
and their interferences in the coherently added cross section of this
reaction. We discuss the importance of $N(1520)$ resonance and show the
sensitivity of the cross section to the hadron nucleus interaction.
\end {abstract}


PACS number(s): 25.40e, 13.30.Eg, 13.60.Le

\vspace{0.5 cm}

The coherent meson production in the nuclear reaction is a potential tool
to investigate the resonance dynamics in the nucleus, as well as the meson
nucleus interaction in the final state. Since the branching ratio of
$\Delta (1232) \to N\pi$ is $\simeq 100\%$ \cite{pdg12}, the coherent pion
production process has been used extensively to investigate the $\Delta$
dynamics in the nucleus \cite{ostw, erwe}.
This
process in $(\gamma,\pi)$ and $(e,e\pi)$ reactions is used to study the
transverse $N \to \Delta$ excitation in the nucleus, where the coherent
pion is produced away from the forward direction \cite{dtks}.
The
forward emission of coherent pion is a probe for the longitudinal $\Delta$
excitation which occur in the pion nuclear reaction \cite{ibpza}.

The coherent pion production is also studied in the proton and ion induced
nuclear reaction \cite{chiba, henn1}. The issue of $\Delta$-peak shift in
the nucleus \cite{contar} is resolved, as it occurs because of the coherent
pion production \cite{chiba, henn1} which is not possible for proton target.
The
coherent pion production in the $(p,n)$ \cite{chiba} and $(^3\mbox{He},t)$
\cite{henn1} reactions on the nucleus is shown to have one to one
correspondence with that in the $\pi^+$ meson nucleus scattering
\cite{das2, kou}.
For
the forward going protons, the coherent pion production in the
$(p,p^\prime)$ reaction can be used to produce $\pi^0$ beam \cite{cnov}
which is analogous to tagged photon beam.

The coherent $\eta$ meson production in the nuclear reaction is another
process which can be used to study the resonance dynamics in the nucleus.
Amongst
the resonances, $N(1535)$ has large decay branching ratio $(42 \%)$ in
the $N\eta$ channel, i.e., $ \Gamma_{N(1535) \to N\eta}
( m=1535 ~\mbox{MeV} ) $ $\approx 63$ MeV \cite{pdg12}. Therefore, this
resonance is considered to study the coherent $\eta$ meson production in
the proton nucleus reaction \cite{oset}.
$N(1535)$
is shown as a sensitive probe to study the non-local effects in the coherent
$\eta$ meson photoproduction reaction \cite{pelm}.
The
importance of $N(1535)$ is elucidated in context of the $\eta$ meson
production in the hadron nucleus reaction \cite{boris}.
In
addition to $N(1535)$, other resonances, e.g., $N(1520)$, $N(1650)$ .....
etc, (whose branching ratio in the $N\eta$ channel is much less than that
of $N(1535)$) and nucleon Born terms are also considered in the study of
$\eta$ meson production in the photonuclear reaction below 1 GeV
\cite{heda}.
The
change in the cross section because of the interference of $N(1520)$ and
$N(1535)$ resonances is described in Ref.~\cite{blsh}.

\begin{figure}[h]
\begin{center}
\centerline {\vbox {
\psfig{figure=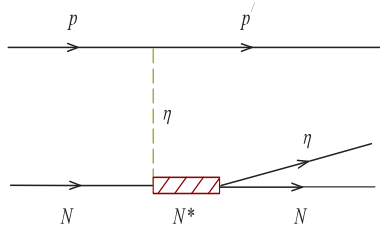,height=3.0 cm,width=5.0 cm}
}}
\caption{
(color online).
Elementary reaction occurring in the nucleus.
}
\label{fg1}
\end{center}
\end{figure}

Sometime back, Alvaredo and Oset studied the coherent $\eta$ meson
production in the $(p,p^\prime)$ reaction on the spin-isospin saturated
nucleus \cite{oset}: $ p + A(gs) \to p^\prime + A(gs) + \eta $.
The
elementary reaction in the nucleus is assumed to proceed as
$ pN \to p^\prime N^* $; $ N^* \to N\eta $ (presented in Fig.~\ref{fg1}).
The resonance $N^*$ considered in the intermediate state is $N(1535)$,
since it has large branching ratio (as mentioned earlier) in the $N\eta$
channel.
This
resonance is produced due to the $\eta$ meson
(a pseudoscalar-isoscalar meson) exchange interaction only, specifically,
for the forward going proton and $\eta$ meson. The contributions from other
meson exchange interactions vanish in this reaction \cite{oset}.
The
projectile excitation in this reaction is null for the spin saturated
nucleus.

\begin{figure}[h]
\begin{center}
\centerline {\vbox {
\psfig{figure=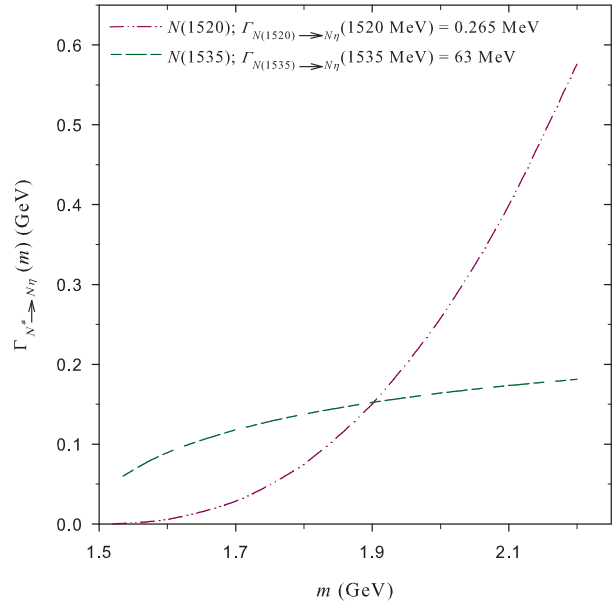,height=08.0 cm,width=08.0 cm}
}}
\caption{
(color online).
Decay width $ \Gamma_{N^* \to N\eta} (m) $ vs resonance mass $m$.
}
\label{fg2}
\end{center}
\end{figure}

It may be argued that though the resonance $N(1520)$ has very small decay
width (at its pole mass) in the $N\eta$ channel, there are enough reasons 
(mentioned below) not to neglect this resonance in the $\eta$ meson
production reaction. \\
(i)
The mass of $N(1520)$ is close to that of $N(1535)$, and therefore there
could be interference effect in the $\eta$ meson production reaction
\cite{blsh}, as quoted above. \\
(ii)
The earlier value for
$ \Gamma_{N(1520) \to N\eta} ( m=1520 ~\mbox{MeV} ) $, as reported in
Ref.~\cite{fiar}, is 0.12 MeV, which corresponds to the coupling constant:
$ f_{\eta NN(1520)} = 6.72 $ \cite{pelm}. According to the recent result:
$ \Gamma_{N(1520) \to N\eta} ( m=1520 ~\mbox{MeV} ) \simeq 0.265 $ MeV
\cite{pdg12}, the value of $f_{\eta NN(1520)}$ is equal to 9.98.
The
later value of $f_{\eta NN(1520)}$ is about 1.5 times larger than its
previous value. Due to this enhancement, the coherent $\eta$ meson
production cross section because of $N(1520)$ is increased by a factor
about 5. \\
(iii)
The decay width $ \Gamma_{N^* \to N\eta} (m) $ varies with the mass $m$
of the resonance $N^*$ \cite{manley} as
\begin{equation}
\Gamma_{N^* \to N\eta} (m) = \Gamma_{N^* \to N\eta} (m_{N^*})
\left [ \frac{\Phi_l (m) }{\Phi_l (m_{N^*})} \right ].
\label{wdth}
\end{equation} 
$m_{N^*}$ in this equation is the pole mass of $N^*$. The value of
$ \Gamma_{N^* \to N\eta} (m_{N^*}) $ is already mentioned for $N(1520)$
and $N(1535)$ resonances.
The
suffix $l$ in the phase-space factor $\Phi_l$ represents the angular
momentum associated with the decay. $\Phi_l$ is given by
$ \Phi_l (m) = \frac{\tilde {k}}{m} B^2_l (\tilde {k}R) $,
where
$\tilde {k}$ is the relative momentum of the decay products
(i.e., $N$ and $\eta$) in their c.m. frame. $ B_l(\tilde {k}R) $ is the
Blatt-Weisskopf barrier-penetration factor, listed in Table \ref{tbb}.
$R$ ($=0.25$ fm) is the interaction radius.
Using
Eq.~(\ref{wdth}), we show in Fig.~\ref{fg2} that the decay probability of
$ N(1520) \to N\eta $ rises sharply over that of $ N(1535) \to N\eta $
with the increase in resonance mass $m$, i.e., $\eta N$ invariant mass.
Therefore,
though the decay width of $ N(1520) \to N\eta $ at the pole mass, as
quoted earlier, is much less than that of $ N(1535) \to N\eta $, the
previous (as shown in Fig.~\ref{fg2}) supersedes the later at higher
values of $m$.
The
steep rise in $ \Gamma_{N(1520) \to N\eta} (m) $ with $m$ can shift the peak
position of the $\eta$ meson production cross section due to $N(1520)$
towards the larger value of $m$.

\begin{table}
\caption{ Blatt-Weisskopf barrier-penetration factor $ B_l(\tilde {k}R) $
          \cite{manley}. }
\centering
\begin{tabular} { c c c }
\hline
$ N^* $    &  $l$   &  $ B^2_l (x = \tilde{k}R) $  \\  [0.5ex]
\hline
$N(1520)$  &   2    &  $ x^4/(9+3x^2+x^4) $        \\
$N(1535)$  &   0    &          1                   \\  [1ex]
\hline          
\end{tabular} 
\label{tbb}
\end{table}

We have elucidated that the cross section due to $N(1520)$ in the $\eta$
meson production reaction could be large. This can change the shape and
magnitude of the coherently added cross section arising due to
nucleon Born terms and resonances.
To
disentangle it, we revisit the coherent $\eta$ meson production in the
$(p,p^\prime)$ reaction on the scalar-isoscalar nucleus where  Born terms,
$N(1520)$ and other resonances of $N\eta$ branching ratio $\geq 4\%$, i.e.,
$N(1535)$, $N(1650)$, $N(1710)$, $N(1720)$, are considered.
In
this reaction, the virtual $\eta$ meson (emitted by projectile) is
elastically scattered to its real state by the nucleus which remains in
its ground state. Since $^4$He does not have excited state, this nucleus
is preferred to study the mechanism of this reaction. In the experiment
both coherent production and breakup will occur, but the coherent channel
can in principle be indentified.

The Lagrangian ${\cal L}$ for the coupling of $\eta$ meson to a particle
depends on their spin and parity \cite{pelm, blsh}. For $\frac{1}{2}^+$
particle, i.e., $ N(940)$ and $N^* \equiv N(1710)$, the form for
${\cal L}$ is
\begin{eqnarray}
{\cal L}_{\eta NN~~} &=& -ig_\eta F_\eta(q^2)
                            {\bar N} \gamma_5 N \eta   \nonumber  \\
{\cal L}_{\eta NN^*} &=& -ig^*_\eta F^*_\eta(q^2)
                              {\bar N^*} \gamma_5 N \eta;  
\label{lag1}
\end{eqnarray}
$g_\eta$ ($\eta NN$ coupling constant) $\simeq 7.93 $ \cite{chol},
and  $g^*_\eta$ ($\eta NN(1710)$ coupling constant) $\simeq 4.26 $. 
For $\frac{1}{2}^-$ resonance $N^*$, i.e., $N(1535)$ and $N(1650)$,
${\cal L}$ is given by
\begin{equation}
{\cal L}_{\eta NN^*} = -ig^*_\eta F^*_\eta(q^2) {\bar N^*} N \eta; 
\label{lag2}
\end{equation}
$ g^*_\eta \simeq 1.86 $ for $N(1535)$ and $ g^*_\eta \simeq 0.67 $
for $N(1650)$.
For $ N(1520) \frac{3}{2}^- $, ${\cal L}$ can be written as 
\begin{equation}
{\cal L}_{\eta NN^*} =  \frac{f^*_\eta}{m_\eta} F^*_\eta(q^2)
                       {\bar N^{*\mu}} \gamma_5 N \partial_\mu \eta;
\label{lag3}
\end{equation}
$ f^*_\eta = 9.98 $.
For $\frac{3}{2}^+$ resonance \cite{das4}, i.e., $ N^* \equiv N(1720) $,
the expression for ${\cal L}$ is
\begin{equation}
{\cal L}_{\eta NN^*} =  \frac{f^*_\eta}{m_\eta} F^*_\eta(q^2)
                       {\bar N^{*\mu}} N \partial_\mu \eta;
\label{lag4}
\end{equation}
$ f^*_\eta = 1.15 $.
The coupling constants are extracted from the measured decay width of the
resonances, i.e., $ N^* \to N\eta $ \cite{pdg12}. $F_\eta$ and $F^*_\eta$
appearing in Lagrangians are the $\eta NN$ and $\eta NN^*$ form factors
respectively \cite{chol}:
$ F_\eta (q^2) = F^*_\eta (q^2)
= \frac{ \Lambda^2_\eta - m^2_\eta }{ \Lambda^2_\eta - q^2 };
 \Lambda_\eta = 1.5 ~ \mbox{GeV}. $

The $T$-matrix for the coherent $\eta$ meson production in the
$(p,p^\prime)$ reaction on a nucleus can be written as
$T_{fi} = T_B+T_{N^*} $, where $T_B$ represents the $T$-matrix for nucleon
Born terms (described later), and the resonance term, i.e., $T_{N^*}$,
is given by
\begin{equation}
T_{N^*} = 
\sum_{N^*}  {\hat \Gamma}_{N^* \to N\eta} \Lambda(S) {\tilde V}_\eta (q)
\int d{\bf r} \chi^{(-)*} ({\bf k}_\eta, {\bf r}) G_{N^*} (m, {\bf r})
\varrho ({\bf r}) \chi^{(-)*} ({\bf k}_{p^\prime}, {\bf r})
\chi^{(+)} ({\bf k}_p, {\bf r}).
\label{tmx1}
\end{equation}
$\varrho ({\bf r})$ in this equation is the matter density distribution
of the nucleus. ${\hat \Gamma}_{N^* \to N\eta}$ denotes the vertex factor
for the decay: $N^* \to N\eta$. The spin $S$ dependent part of $N^*$
propagator, i.e., $\Lambda(S)$, is expressed in Table-\ref{tbs}.

\begin{table}
\caption{ $\Lambda(S)$ for spin $S=$ $\frac{1}{2}$ and
$\frac{3}{2}$ fermions. }
\centering
\begin{tabular} { c c }
\hline
Spin($S$)      &  $\Lambda(S)$    \\  [0.5ex]
\hline
$\frac{1}{2}$  &  $\{ \not k + m_{N^*} \} $  \\
$\frac{3}{2}$  &
$ \{ \not k + m_{N^*} \}
\left [ g^\mu_\nu  -\frac{\gamma^\mu \gamma_\nu}{3}
-\frac{ \gamma^\mu k_\nu - \gamma^\nu k_\mu }{ 3m_{N^*} }
-\frac{2k^\mu k_\nu}{3m^2_{N^*}}
\right  ]$   \\  [1ex]
\hline
\end{tabular} 
\label{tbs}
\end{table}

${\tilde V}_\eta (q)$ in above equation represents the $\eta$
meson exchange interaction between the beam proton and the nucleon in the
nucleus (see Fig.~1):
$ {\tilde V}_\eta (q) = {\hat \Gamma}_{\eta NN^*} {\tilde G}_\eta (q^2)
                   {\hat \Gamma}_{\eta NN}.  $
The ${\hat \Gamma}$s are described by the Lagrangians given in
Eqs.~(\ref {lag1}) - (\ref{lag4}).
${\tilde G}_\eta (q^2)$ denotes the virtual $\eta$ meson propagator,
given by $ {\tilde G}_\eta (q^2) = - \frac{1}{m^2_\eta - q^2} $.

The distorted wave functions for proton and $\eta$ meson, denoted by
$\chi$s in Eq.~(\ref {tmx1}), are evaluated by using Glauber model
\cite{das5}. For the beam proton $p$, it can be written as
\begin{equation}
\chi^{(+)} ({\bf k}_p, {\bf r}) = e^{i {\bf k}_p. {\bf r} }
exp[ -\frac{i}{v_p} \int^z_{-\infty} dz^\prime V_{Op} ({\bf b}, z^\prime) ].
\label{dwptn}
\end{equation}
For outgoing particles, i.e., $p^\prime$ and $\eta$ meson, $\chi$ is
given by
\begin{equation}
\chi^{(-)^*} ({\bf k}_{p^\prime (\eta)}, {\bf r})
= e^{-i {\bf k}_{p^\prime (\eta)}. {\bf r} }
exp[ -\frac{i}{v_{p^\prime (\eta)}}
\int^{+\infty}_z dz^\prime V_{O{p^\prime (\eta)}} ({\bf b}, z^\prime) ].
\label{dwpet}
\end{equation}
$v_X$ is the velocity of the particle $X$ which is either proton or $\eta$
meson appearing in Eqs.~(\ref{dwptn}) and (\ref{dwpet}). $V_{OX}$ denotes
the particle $X$ nucleus optical potential. This potential, in fact,
describes the initial and final state interactions.

The proton nucleus optical potential $V_{Op(p^\prime)} ({\bf r})$
in Eqs.~(\ref{dwptn}) and (\ref{dwpet}) is calculated using the
$ ``t\varrho ({\bf r})" $ approximation \cite{das5}, i.e.,
\begin{equation}
V_{Op} ({\bf r})
= -\frac{v_p}{2} [i+\alpha_{pN}] \sigma^{pN}_t \varrho ({\bf r}),
\label{opts}
\end{equation}
where $\alpha_{pN}$ denotes the ratio of
the real to imaginary part of the proton nucleon scattering amplitude
$f_{pN}$. $\sigma^{pN}_t$ represents the corresponding total cross section.
To evaluate this potential, we use the energy dependent experimentally
determined values for $\alpha_{pN}$ and $\sigma^{pN}_t$ \cite{nndt}.

The $\eta$ meson nucleus optical potential $V_{O\eta}$ ({\bf r}) in
Eq.~(\ref{dwpet}), following Alvaredo and Oset \cite{oset}, is
evaluated from the $\eta$ meson self-energy $\Pi_\eta$ ({\bf r}) in the
nucleus:
\begin{eqnarray}
\Pi_\eta ({\bf r}) = 2E_\eta V_{O\eta} ({\bf r}) =
\sum_{N^*} |C(N^*)|^2
\frac{ \varrho ({\bf r}) }
{ m - m_{N^*} + \frac{i}{2}\Gamma_{N^*}(m) - V_{ON^*} ({\bf r})
                                           + V_{ON} ({\bf r}) }.
\label{opet}
\end{eqnarray}
The prefactor $C(N^*)$ depends on the resonance $N^*$ used to calculate
$\Pi_\eta ({\bf r})$.
$\Gamma_{N^*}(m)$
represents the total width of $N^*$ for its mass equal to $m$. It is
composed of partial widths of $N^*$ decaying into various channels,
listed duly along with the physical parameters in Ref~\cite{pdg12}.
The
resonance mass $m$ dependence of these widths are worked out following
Eq.~(\ref{wdth}). Values of $\Gamma_{N^*}(m)$ at the pole mass, i.e.,
$m=m_{N^*}$, are given in Table-\ref{tbw}.
$V_{ON^*}$
is the $N^*$ nucleus interacting potential, described latter.
The nucleon potential energy in the nucleus is taken as
$ V_{ON} ({\bf r}) = -50 \varrho ({\bf r}) / \varrho (0) $ MeV \cite{oset}.
$\Pi_\eta ({\bf r})$
arising due to nucleon-hole pair is evaluated following that for $\pi^0$
meson, see page no.~157 in Ref.~\cite{erwe}.

\begin{table}
\caption{ Resonance width $ \Gamma_{N^*} (m_{N^*}) $ at pole mass
$m_{N^*}$ in MeV \cite{pdg12}. }
\centering
\begin{tabular} { c c }
\hline
Resonance $N^*$ & $ \Gamma_{N^*} (m_{N^*}) $  \\  [0.5ex]
\hline
$N(1520)$       &    115                      \\
$N(1535)$       &    150                      \\
$N(1650)$       &    150                      \\
$N(1710)$       &    100                      \\
$N(1720)$       &    250                      \\  [1ex]
\hline
\end{tabular} 
\label{tbw}
\end{table}

The scalar part of the resonance propagator,
denoted by $ G_{N^*} (m, {\bf r}) $ in Eq.~(\ref{tmx1}), is given by
\begin{equation}
G_{N^*} (m, {\bf r}) = \frac{1}
{m^2-m^2_{N^*} + im_{N^*}\Gamma_{N^*}(m) - 2E_{N^*}V_{ON^*}({\bf r})},
\label{nspr}
\end{equation}
where $E_{N^*}$ denotes the energy of $N^*$.

$V_{ON^*}({\bf r})$ in Eqs.~(\ref{opet}) and (\ref{nspr}), which describes
the $N^*$ nucleus interaction, is also evaluated by using the
$ ``t\varrho ({\bf r})" $ approximation, as given in Eq.~(\ref{opts}).
In
this case, the measured values of the $N^*$ nucleon scattering parameters,
i.e., $\alpha_{N^*N}$ and $\sigma^{N^*N}_t$, are not available. To estimate
them, we take $ \alpha_{N^*N} \simeq \alpha_{NN} $ and
$ \sigma^{N^*N}_{el} \simeq \sigma^{NN}_{el} $ since the elastic scattering
dynamics of $N^*$ can be assumed not much different from that of a nucleon.
For
the reactive part of $\sigma^{N^*N}_t$, we consider the dynamics of $N^*$ is
same as that of a nucleon at its kinetic energy enhanced by $\Delta m$,
i.e.,
$ \sigma_r^{N^*N} (T_{N^*N}) \approx \sigma_r^{NN} (T_{N^*N}+\Delta m) $
\cite{jakl}.
Here, $\Delta m$ is the mass difference between the resonance and
nucleon. $T_{N^*N}$ is the total kinetic energy in the $N^*N$ center
of mass system \cite{jakl}.

The $\eta$ meson emitted because of the nucleon Born terms is addressed by
$T_B$ above Eq.~(\ref{tmx1}). It is similar to the expression appearing
in this equation except the interaction vertices and propagator of $N^*$
to be replaced by those of nucleon.
The
previous is described by ${\cal L}_{\eta NN}$ in Eq.~(\ref{lag1}) and the
later, i.e., propagators $G_N$, are discussed in Ref.~\cite{ostw}. However,
the calculated cross section due to Born terms, as shown later, is negligibly
small.

\begin{figure}[h]
\begin{center}
\centerline {\vbox {
\psfig{figure=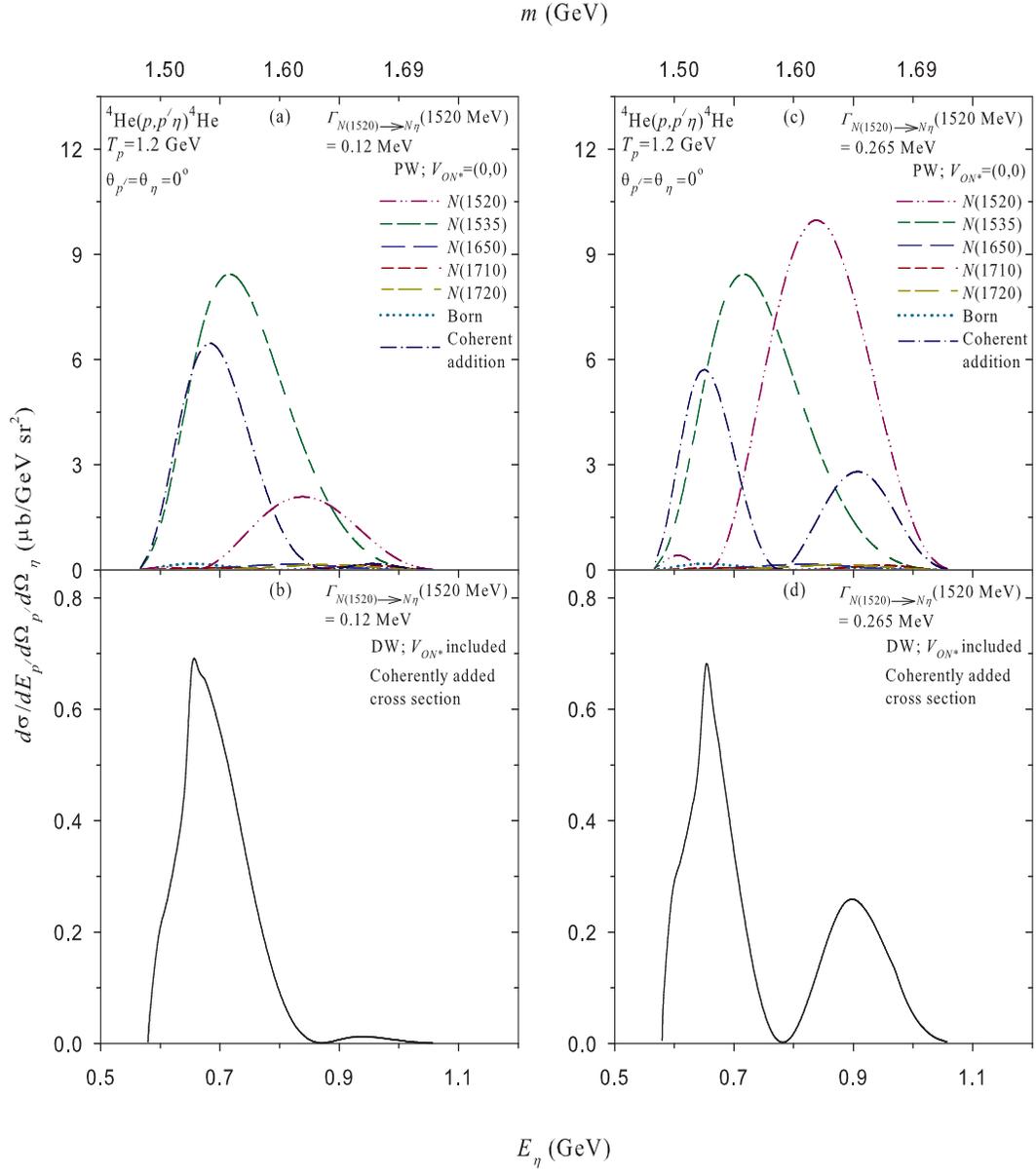,height=16.0 cm,width=14.0 cm}
}}
\caption{
(color online).
Cross sections for the coherent $\eta$ meson production at $T_p=1.2$
GeV. $m$ is the resonance mass corresponding to the $\eta$ meson energy
$E_\eta$ (see text).
}
\label{fg3}
\end{center}
\end{figure}

\begin{figure}[h]
\begin{center}
\centerline {\vbox {
\psfig{figure=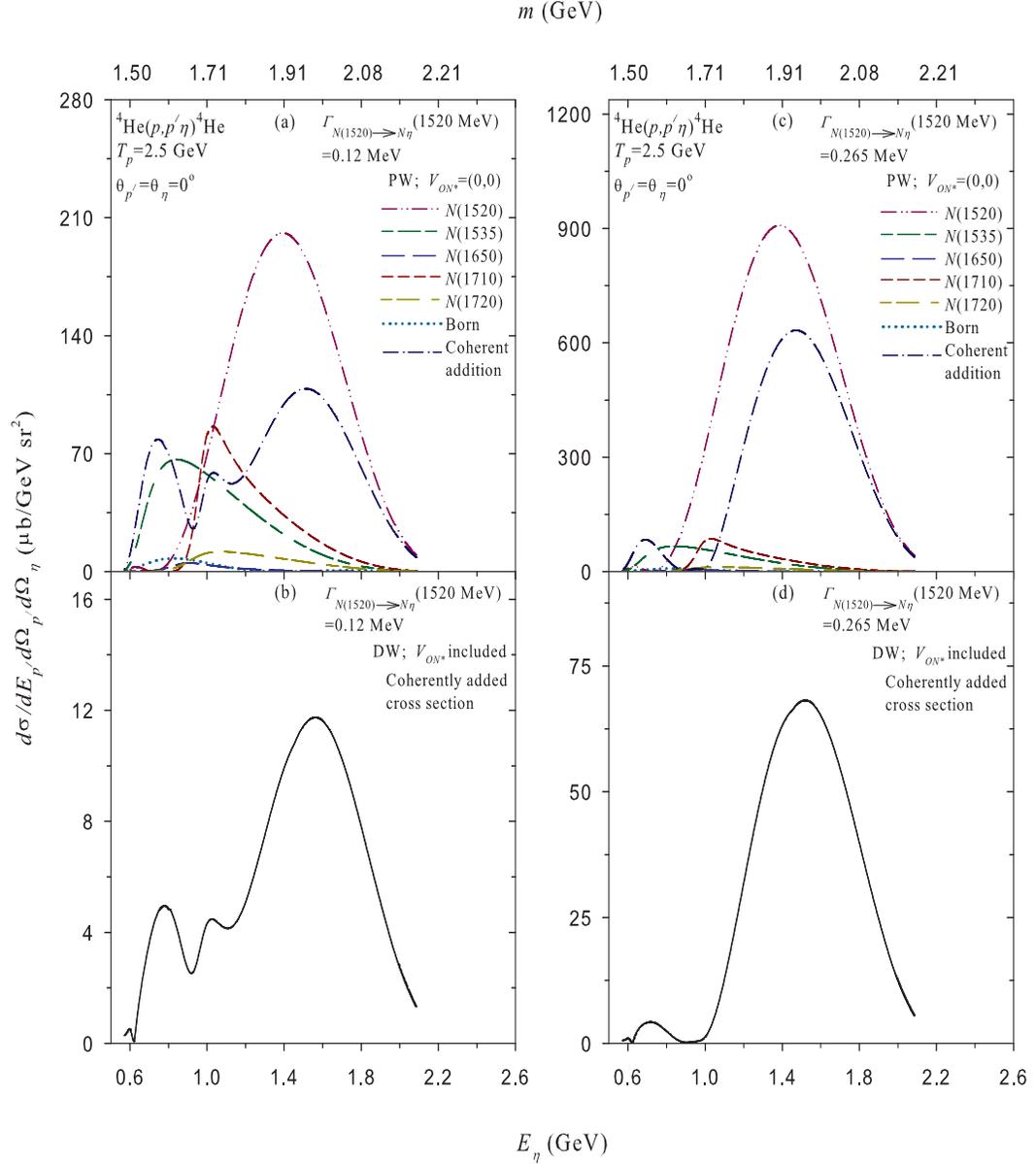,height=16.0 cm,width=14.0 cm}
}}
\caption{
(color online).
Same as Fig.~\ref{fg3} but for the beam energy $T_p=2.5$ GeV (see text). 
}
\label{fg4}
\end{center}
\end{figure}

We calculate the differential cross section, i.e.,
$ \frac { d\sigma } { dE_{p^\prime} d\Omega_{p^\prime} d\Omega_\eta} $,
for the forward going coherent $\eta$ meson energy $E_\eta$ distribution
in the $(p,p^\prime)$ reaction on $^4$He nucleus.
The calculated results are presented in Figs.~\ref{fg3} and \ref{fg4}. On
the upper x-axis of these figures, we mention the resonance mass $m$
corresponding to $E_\eta$.
The
spatial density distribution $ \varrho ({\bf r}) $ of this nucleus is
$\varrho ({\bf r}) =
\varrho_0 \frac { 1+w(r/c)^2 }{ 1+e^{(r-c)/z} }; ~w=0.445, ~c=1.008
~\mbox{fm}, ~z=0.327 ~\mbox{fm}$ \cite{andt}.
We
discuss the contribution of each resonance and nucleon Born terms
(including their interferences) to the cross section of this reaction.
We focus on two aspects of $N(1520)$ which can change the $\eta$ meson
production cross section considerably:
(i) measured decay width
$ \Gamma_{N(1520) \to N\eta} (m=1520 ~\mbox{MeV}) $ and
(ii) decay probability $ N(1520) \to N\eta $ specifically at high energy.

The calculated plane wave results at $T_p=1.2$ GeV are illustrated in
Fig.~\ref{fg3}(a), where the $N^*$ nucleus interaction is not included,
i.e., $V_{ON^*}=(0,0)$. The cross sections arising due to each
resonance and Born terms along with their coherent contribution are shown
in this figure. For $N(1520)$ resonance, the measured value of its decay 
width $ \Gamma_{N(1520) \to N\eta} (m=1520 ~\mbox{MeV}) $ is taken equal
to 0.12 MeV (earlier value).
This
figure elucidates the cross section due to $N(1535)$, shown by
short-long-short dash curve, is the largest. The dot-dot-dash curve
represents the second largest cross section (which is $24.7 \%$ of the
previous at the peak) arises because of $N(1520)$.
The
peak cross section due to it appears at higher value of $E_\eta$
(which corresponds to larger $m$) compared to that because of $N(1535)$.
This occurs, as described in Fig.~\ref{fg2}, due to the sharp rise in
$N(1520) \to N\eta$ decay probability with $m$.
Fig.~\ref{fg3}(a)
also shows that the cross sections arising because of other resonances
and Born terms are negligibly small.
The
coherently added cross section (dot-dash curve) shows that the dominant
contribution to it arises due to $N(1535)$. The effect of interference in
the cross section is distinctly visible in this figure.
In
Fig.~\ref{fg3}(b), we present the distorted wave results where $V_{ON^*}$
is also incorporated. It shows that the cross section is reduced
drastically, i.e., by a factor of 9.6 at the peak, and the peak position
is shifted by 32 MeV towards the lower value of $E_\eta$ due to the
inclusion of distortions (both initial and final states) and $V_{ON^*}$.
These
are the features usually reported in the low energy $\eta$ meson production
reaction.

As stated earlier, the calculated cross section due to $N(1520)$ can go
up by a factor of $\sim 5$ because of the use of the latest measured value
of $ \Gamma_{N(1520) \to N\eta} (m=1520 ~\mbox{MeV}) $, i.e., 0.265 MeV. 
This can change the $\eta$ meson energy $E_\eta$ distribution of the
coherently added cross section. We illustrate those in Fig.~\ref{fg3}(c).
It
is remarkable that the cross section due to $N(1520)$ for
$ \Gamma_{N(1520) \to N\eta} (m=1520 ~\mbox{MeV}) = 0.265 $ MeV
(dot-dot-dash curve) is comparable with that because of $N(1535)$
(short-long-short curve). In fact, the previous is about $18\%$ larger than
the later.
Due
to this, an additional peak in the spectrum of the coherently added cross
section (dot-dash curve) appears close to the peak arising because of
$N(1520)$.
The
magnitude of the cross section due to the inclusion of distortions and
$V_{ON^*}$, as shown in Fig.~\ref{fg3}(d), is reduced drastically, but
the change in the shape of spectrum due to them is insignificant.
Two
peaks are distinctly visible in this figure.

The decay probability of $ N(1520) \to N\eta $, as shown in
Fig.~\ref{fg2}, rises sharply at higher values of the resonance mass
$m$. Therefore, this resonance could be more significant in the high
energy $\eta$ meson production reaction.
To
investigate it, we calculate the plane wave (without $V_{ON^*}$ included)
cross sections at $T_p=2.5$ GeV (taking the earlier value of
$ \Gamma_{N(1520) \to N\eta} (m=1520 ~\mbox{MeV}) $, i.e., 0.12 MeV), and
present those in Fig.~\ref{fg4}(a).
The
cross sections arising because of each resonance and Born terms along with
the coherently added cross section are distinctly visible in this figure.
Compare
to the spectra presented in Fig.~\ref{fg3}(a), there are considerable
changes in those at $T_p=2.5$ GeV which are noticeable in Fig.~\ref{fg4}(a).
Three distinct peaks in the later figure arise due to $N(1520)$
(dot-dot-dash curve), $N(1535)$ (short-long-short dash curve) and $N(1710)$
(short dash curve).
Amongst
them, the cross section because of $N(1520)$ is the largest. The cross
sections due to resonances other than these three resonances and
Born terms are insignificant.
The
separation between the peaks of the cross sections due to $N(1520)$ and
$N(1535)$ at $T_p=2.5$ GeV is larger than that found at $T_p=1.2$ GeV
(see Figs. in \ref{fg3}(a) and \ref{fg4}(a)).
This
occurs, as illustrated in Fig.~\ref{fg2}, because of the sharp increase in
$N(1520) \to N\eta$ decay probability with $m$.

Because of the large $\eta$ meson production cross section due to $N(1520)$
at $T_p=2.5$ GeV (even for 
$ \Gamma_{N(1520) \to N\eta} (m=1520 ~\mbox{MeV}) = 0.12 $ MeV), the
shape of the $\eta$ meson energy $E_\eta$ distribution spectrum of the
coherently added cross section is significantly different from that
calculated at $T_p=1.2$ GeV (see dot-dash curves in Fig.~\ref{fg3}(a) and
Fig.~\ref{fg4}(a)).
We show the distorted wave ($V_{ON^*}$ included) result at $T_p=2.5$ GeV
in Fig.~\ref{fg4}(b).
Both
Figs.~\ref{fg4}(a) and ~\ref{fg4}(b) show the coherently added cross
sections possess multiple peaks appearing in the peak regions of the
cross sections due to $N(1520)$, $N(1535)$ and $N(1710)$ resonances.
In addition, these figures elucidate that the largest peak arises in the
$N(1520)$ excitation region.
These
results are unlike to those presented in Figs.~\ref{fg3}(a) and \ref{fg3}(b).

The calculated plane wave, without $V_{ON^*}$ included, cross section
at $T_p=2.5$ GeV due to $N(1520)$ for the latest value of
$ \Gamma_{N(1520) \to N\eta} (m=1520 ~\mbox{MeV}) $, i.e., 0.265 MeV,
is presented by the dot-dot-dash curve in Fig.~\ref{fg4}(c).
Along
with it, the cross sections because of other resonances and Born terms are
also presented for comparison.
Due
to the increase in $ \Gamma_{N(1520) \to N\eta} (m=1520 ~\mbox{MeV}) $,
in addition to large $N(1520) \to N\eta$ decay probability at higher energy,
the $\eta$ meson production (as shown in this figure) dominantly occurs
because of the $N(1520)$ resonance.
In fact, the peak cross section due to it is distinctly largest.
The
coherently added cross section (dot dash curve in Fig.~\ref{fg4}(c)) shows
a small peak close to that because of $N(1535)$ and a large peak
distinctly visible near to that due to $N(1520)$.
We
illustrate the distorted wave ($V_{ON^*}$ included) result for it in
Fig.~\ref{fg4}(d).
The
dominant contribution to the coherently added cross section, as shown in
Figs.~\ref{fg4}(c) and \ref{fg4}(d), distinctly arises because of the
resonance $N(1520)$.

We have calculated the differential cross sections for the coherent $\eta$
meson energy $E_\eta$ distribution in the $(p,p^\prime)$ reaction on
$^4$He nucleus.
At
lower beam energy, i.e., 1.2 GeV, the cross section because of $N(1535)$ is
distinctly largest if we consider the earlier value of the decay width
$ \Gamma_{N(1520) \to N\eta} (m=1520 ~\mbox{MeV}) $, i.e., 0.12 MeV. 
The
cross section due to $N(1520)$ is drastically increased because of the
use of the latest value of
$ \Gamma_{N(1520) \to N\eta} (m=1520 ~\mbox{MeV}) $, i.e., 0.265 MeV.
Due
to this, an additional peak appears in the shape of the coherently added
cross section.
At
higher beam energy, i.e., 2.5 GeV, the peak cross section because of
$N(1520)$, even for
$ \Gamma_{N(1520) \to N\eta} (m=1520 ~\mbox{MeV}) = 0.12 $ MeV, supersedes
that due to $N(1535)$.
The shift of $N(1520)$-peak towards the higher $\eta$ meson energy (which
corresponds to larger resonance mass) depends on the beam energy.
This
occurs because of the sharp increase in $ N(1520) \to N\eta $ decay
probability with the resonance mass.
The
cross section due to $N(1520)$ is further increased by a factor $\sim 5$
because of the increase in
$ \Gamma_{N(1520) \to N\eta} (m=1520 ~\mbox{MeV}) $ from 0.12 MeV to
0.265 MeV.
The
coherently added cross section shows the contribution from $N(1520)$ is
distinctly dominant amongst the resonances and Born terms at higher energy.
These
features are unlike those reported in the previous studies where $N(1535)$
is shown to contribute dominantly in the $\eta$ meson production reaction.

The author gratefully acknowledges the referee for giving valuable
comments on this work.

\newpage

{\bf Figure Captions}
\begin{enumerate}

\item
(color online).
Elementary reaction occurring in the nucleus.

\item
(color online).
Decay width $ \Gamma_{N^* \to N\eta} (m) $ vs. resonance mass $m$.

\item
(color online).
Cross sections for the coherent $\eta$ meson production at $T_p=1.2$
GeV. $m$ is the resonance mass corresponding to the $\eta$ meson energy
$E_\eta$ (see text).

\item
(color online).
Same as Fig.~\ref{fg3} but for the beam energy $T_p=2.5$ GeV (see text).

\end{enumerate}

\newpage

\newpage

\newpage

\newpage

\end{document}